\documentclass[sigconf]{acmart}

\usepackage{hyperref}
\usepackage{url}
\usepackage{subfigure}
\usepackage{graphicx}
\usepackage{multirow}
\usepackage{algorithm}
\usepackage{algorithmic}

\AtBeginDocument{%
  \providecommand\BibTeX{{%
    \normalfont B\kern-0.5em{\scshape i\kern-0.25em b}\kern-0.8em\TeX}}}

\begin{document}

\title{Learning Preferences and Demands in Visual Recommendation}

\author{Qiang Liu}
\affiliation{%
  \institution{RealAI \& Tsinghua University}
  \city{Beijing}
  \country{China}}
\email{qiang.liu@realai.ai}

\author{Shu Wu}
\affiliation{%
  \institution{Institute of Automation, Chinese Academy of Sciences}
  \city{Beijing}
  \country{China}}
\email{shu.wu@nlpr.ia.ac.cn}

\author{Liang Wang}
\affiliation{%
  \institution{Institute of Automation, Chinese Academy of Sciences}
  \city{Beijing}
  \country{China}}
\email{liang.wang@nlpr.ia.ac.cn}

\renewcommand{\shortauthors}{Liu, et al.}

\begin{abstract}
Visual information is an important factor in recommender systems, in which users' selections consist of two components: \emph{preferences} and \emph{demands}. Some studies has been done for modeling users' preferences in visual recommendation. However, conventional methods models items in a common visual feature space, which may fail in capturing \emph{styles} of items. We propose a DeepStyle method for learning style features of items. DeepStyle eliminates the categorical information of items, which is dominant in the original visual feature space, based on a Convolutional Neural Networks (CNN) architecture. For modeling users' demands on different categories of items, the problem can be formulated as recommendation with contextual and sequential information. To solve this problem, we propose a Context-Aware Gated Recurrent Unit (CA-GRU) method, which can capture sequential and contextual information simultaneously. Furthermore, the aggregation of prediction on preferences and demands, i.e., prediction generated by DeepStyle and CA-GRU, can model users' selection behaviors more completely. Experiments conducted on real-world datasets illustrates the effectiveness of our proposed methods in visual recommendation.
\end{abstract}





\maketitle

\section{Introduction}

Nowadays, it is important to sensing and understanding what users prefer and need in recommender systems, which have been fundamental components of various applications. People always say ``Seeing is believing." \emph{Visual} information plays an important role in understanding user behaviors, especially in domains such as cloths, jewelries, house decorations and so on. It is crucial to investigate the visual dimensions of users' preferences and items' characteristics for better personalized recommendation.

\begin{figure}[!tb]
\centering
\includegraphics[width=1\linewidth]{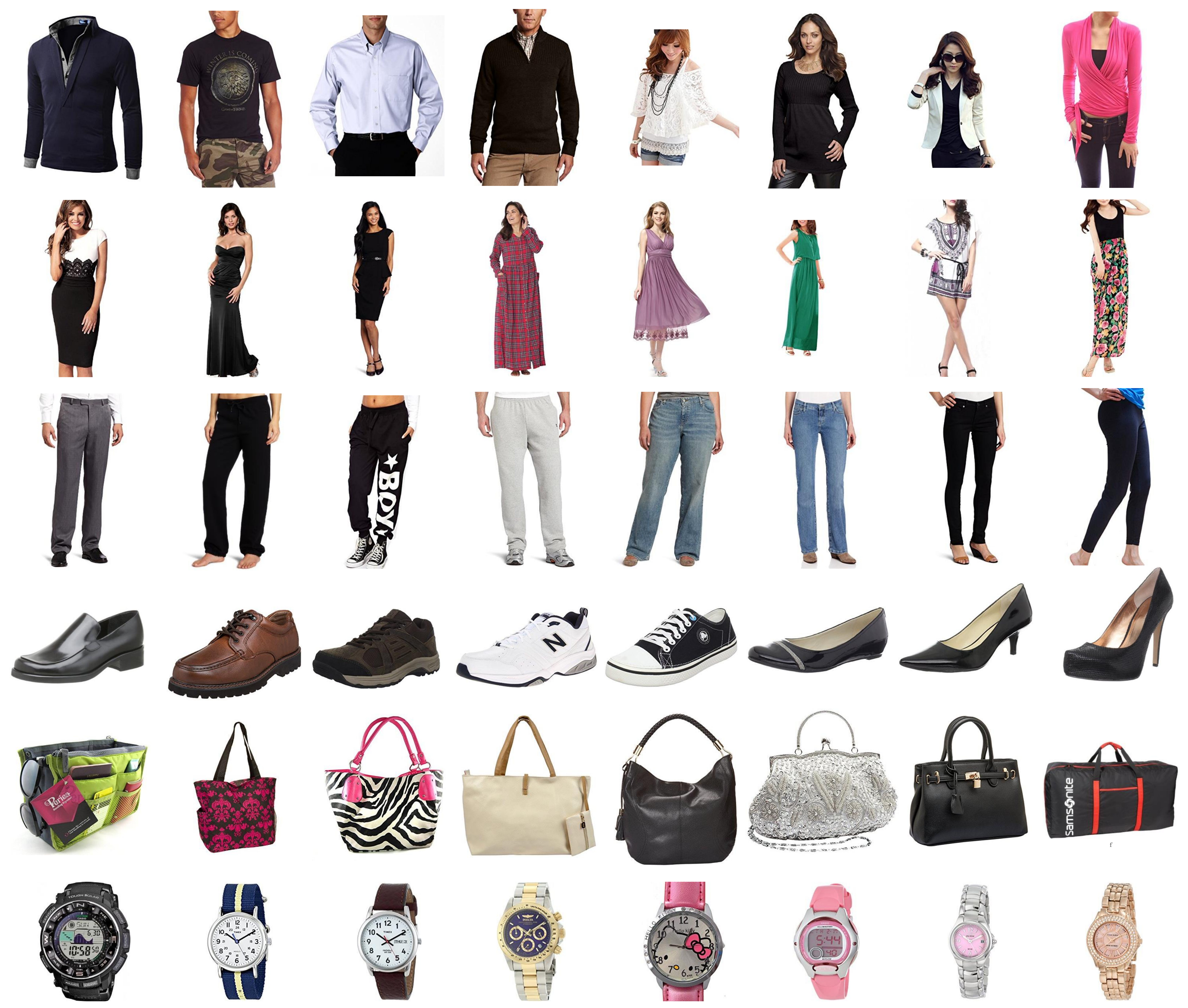}
\caption{Part of the clustering results of items in the Clothing subset of the Amazon dataset \cite{mcauley2015inferring,mcauley2015image} measured by CNN visual features \cite{jia2014caffe,krizhevsky2012imagenet}. One row is a cluster. We can observe that, each category of items are assigned to one cluster. Male and female items are not distinguished. Different styles of clothing are neither distinguished.}
\label{fig:origin}
\end{figure}

Recently, some studies have been done on investigating visual features for user modeling, including cloth matching \cite{hu2015collaborative,mcauley2015image} and visual recommendation \cite{he2016ups,he2016vbpr}. However, these methods models items in a common visual feature space, which may fail to capture different styles of items. In Figure \ref{fig:origin}, we cluster items in the clothing subset of the Amazon dataset\footnote{http://jmcauley.ucsd.edu/data/amazon/} \cite{mcauley2015inferring,mcauley2015image}. The visual features used here are the Convolutional Neural Networks (CNN) visual features extracted from the Caffe reference model\footnote{bvlc\_reference\_caffenet from caffe.berkeleyvision.org} \cite{jia2014caffe,krizhevsky2012imagenet}, which have been also used in several existing works \cite{he2016sherlock,he2016ups,he2016vbpr,mcauley2015image}. We can observe that, each category (e.g., ups, dresses, pants, shoes, bags and watches) of items are assigned to one cluster. It is obvious that, different styles (e.g., casual, athletic and formal) of items can not be distinguished in the figure, even if the male and female styles. Items with similar styles are usually bought together, but they are not similar in the visual feature space. Thus, it is hard for a recommender to make reliable prediction in such feature space. For example, the similarity between suit pants and leather shoes is much small than the similarity between suit pants and jeans. However, suit pants and leather shoes are usually bought together by the same user. Thus, we need to investigate the \emph{styles} of items, and eliminate the characteristics of categories from representations of items. Accordingly, we assume that:
\begin{equation} \label{eq:ass_style}
item = style + category~.
\end{equation}
Based on the assumption in Equation \ref{eq:ass_style}, we propose a novel method called \textbf{DeepStyle}. In DeepStyle, images of items are feeded into a deep CNN model. For each item, on the output layer of CNN, we subtract a latent representation of the corresponding category from the visual feature vector generated by CNN, and thus, we obtain the style features of items. Then, we incorporate style features with the widely-used Bayesian Personalized Ranking (BPR) \cite{rendle2009bpr} for personalized recommendation.

Moreover, as we eliminate the characteristics of categories from representations of items, we need to find another way to take usage of the categorical information, which is also important for modeling user behaviors. Here, we extract two components in users' selection behaviors (e.g., buying a pair of jeans): \emph{preferences} and \emph{demands}. Usually, a user may have a category of items that he or she needs (demands), then the user selects one that he or she likes belonging to the category (preferences). Thus, we can achieve an assumption:
\begin{equation} \label{eq:ass_demand}
selection = preference + demand~.
\end{equation}
Usually, demands lead to categories (e.g., coats, shoes, sandals, shirts and dresses), while preferences lead to styles (e.g., casual, formal, fashionable and conservative). For example, when the weather becomes cold, you will need a coat (demands) of the casual style (preferences). Preferences on styles of a user are comparative permanent and stable in a relatively long period, which can be modeled by DeepStyle introduced above. Demands on categories are temporary, and depend on situations and what a user has bought before. For example, when the weather is hot, you will need sandals instead of boots. And when you have bought lots of shoes, you probably do not need another pair. Thus, predicting the category a user needs becomes a recommendation problem with contextual and sequential information.

Context-aware recommendation \cite{liu2015cot,rendle2011fast,shi2014cars} and sequential recommendation \cite{rendle2010factorizing,wang2015learning,yu2016dream} are both extensively studied. In our recent work \cite{liu2016context}, the problem of context-aware sequential recommendation has been addressed, and a model called Context-Aware Recurrent Neural Networks (CA-RNN) is proposed. CA-RNN captures sequential and contextual information simultaneously in a Recurrent Neural Networks (RNN) architecture. It adjusts matrices in the conventional RNN formulation to variety of contexts. However, models with conventional RNN architecture may occur the vanishing or exploding gradients problem \cite{bengio1994learning}, and assigning each context with a matrix requires too many parameters. Accordingly, we incorporate the Gated Recurrent Unit (GRU) \cite{chung2014empirical,chung2015gated} architecture, and propose a novel method, namely Context-Aware Gated Recurrent Unit (\textbf{CA-GRU}). In the formulation of CA-GRU, we incorporate context-aware gates for adjusting to different contexts. Comparing with context-aware matrices in CA-RNN, context-aware gate vectors in CA-GRU require much less parameters. And with the GRU architecture, the vanishing or exploding gradients problem can be relieved.

Finally, the prediction on preferences and demands, i.e., the prediction generated by DeepStyle and CA-GRU, can be aggregated, for better prediction on users' selections, and promoting the performance in visual recommendation. The main contributions of this work are listed as follows:
\begin{itemize}
\item
We address two components in users' selection behaviors: preferences and demands, leading to styles and categories of items respectively.

\item
We propose DeepStyle for learning style features of items, and sensing users' preferences.

\item
We propose CA-GRU for context-aware sequential recommendation, which can be applied for better predicting users' demands on different categories of items.

\item
Experiments conducted on real-world datasets demonstrates the effectiveness of our proposed methods in visual recommendation.

\end{itemize}

The rest of the paper is organized as follows. In section 2, we review some related works. Section 3 details our proposed DeepStyle and CA-GRU for learning users' preferences and demands respectively. In section 4, we report and analyze our experimental results. Section 5 concludes our work and discusses future research.

\section{Related Works}

In this section, we review some related works on visual recommendation, context-aware recommendation and sequential recommendation.

\subsection{Visual Recommendation}

Matrix Factorization (MF) \cite{koren2009matrix} and Bayesian Personalized Ranking (BPR) \cite{rendle2009bpr} have become the state-of-the-art approaches to recommender systems. Based on these methods, some extended methods are focusing on incorporating visual features for better user modeling. And the importance of utilizing visual features of items in recommender systems has been stressed and proved \cite{di2014picture,goswami2011study,he2016sherlock}. Functional Pairwise Interaction Tensor Factorization (FPITF) \cite{hu2015collaborative} predicts the matching of clothes in outfits with tensor factorization. Personalized matching of items based on visual features has also been investigated \cite{mcauley2015image}. Deep CNNs have been incorporated to model visual features of clothing according to dyadic co-occurrences \cite{veit2015learning}. Collaborative Knowledge-base Embedding (CKE) \cite{zhang2016collaborative} incorporates knowledge-bases, including visual knowledge, for recommender systems. Visual Bayesian Personalized Ranking (VBPR) \cite{he2016vbpr} extends the framework of BPR, and incorporates visual features for promoting the recommendation of items in implicit feedback scenarios. VBPR is further extended with dynamic dimensions to model the visual evolution of fashion trends in visual recommendation \cite{he2016ups}. The impact of categorical information on items' styles has been considered in Sparse Hierarchical Embeddings (Sherlock) \cite{he2016sherlock}. In Sherlock, the embedding matrices for transferring visual features to style features vary among different categories of items. However, Sherlock requires a prior category tree. Moreover, one embedding matrix for each category leads to very large amount of parameters to be learned, although a sparse operation on the category tree is performed.

\subsection{Context-aware and Sequential Recommendation}

Context-aware recommendation \cite{adomavicius2011context} is a major topic in recommender systems. Multi-verse recommendation \cite{karatzoglou2010multiverse} uses tensor factorization to model $n$-dimensional contextual information. As a extended model of tensor factorization, Factorization Machine (FM) can model a wide variety of contextual information by specifying contextual information as the input dimensions and provide context-aware predictions \cite{rendle2011fast}. The Tensor Factorization for MAP maximization (TFMAP) model \cite{shi2012tfmap} uses tensor factorization and Mean Average Precision (MAP) objective to model implicit feedback data with contextual information. Recently, CARS2 \cite{shi2014cars} and Contextual Operating Tensor (COT) \cite{liu2015cot,wu2016contextual,wu2017contextual} models represent the common semantic effects of contexts as contextual operating tensor and represents contexts as latent vectors. And the Hierarchical Interaction Representation (HIR) model \cite{liu2015collaborative,wu2017hierarchical} generates the interaction representation via tensor multiplication which can be applied in context-aware recommender system.

Methods based on Markov assumption are most widely-used models for sequential recommendation \cite{yang2010personalizing}. Via factorization of the probability transition matrices, Factorizing Personalized Markov Chain (FPMC) \cite{rendle2010factorizing} can provide better personalized prediction for sequential recommendation. Hierarchical Representation Model (HRM) \cite{wang2015learning} learns the representation of behaviors in the last transaction and predicts behaviors for the next transaction. Recently, Recurrent Neural Networks (RNN) have achieved the state-of-the-art performance in sequential prediction. RNN based models have been applied successfully in many different scenarios, such as sentence modeling \cite{mikolov2010recurrent}, sequential click prediction \cite{zhang2014sequential}, location prediction \cite{liu2016strnn}, next basket recommendation \cite{yu2016dream} and session-based recommendation \cite{hidasi2016session}. Back Propagation Through Time (BPTT) \cite{rumelhart1988learning} is usually employed to learn parameters of these RNN based models. Recently, Convolutional Neural Networks (CNN) have also be incorporated for sequential prediction \cite{tang2018personalized,yuan2019simple,wang2019towards}.

Recently, the problem of context-aware sequential recommendation has been addressed for capturing sequential and contextual information simultaneously \cite{liu2016context,wu2017context}. To solve this problem, Context-Aware Recurrent Neural Networks (CA-RNN) is proposed based on a RNN architecture. CA-RNN focuses on two types of contexts: input contexts and transition contexts. Input contexts denote situations where users conduct behaviors, while transition contexts mean time intervals between adjacent behaviors in sequences. CA-RNN adjusts matrices in the conventional RNN formulation to variety of contexts. However, models with the conventional RNN architecture may occur the vanishing or exploding gradients problem \cite{bengio1994learning}. Moreover, assigning each context with a matrix in CA-RNN requires too many parameters, and may result in overfitting.

\section{Proposed Methods}

In this section, we detail our proposed methods. First, we give the notations in this work. Then, we introduce DeepStyle and CA-GRU for learning users' preferences and demands respectively. Finally, we present the aggregation of DeepStyle and CA-GRU for better predicting users' selections in visual recommendation.

\subsection{Notations}

In this work, we focus on predicting users' implicit feedbacks, i.e., users' selections, on items. We have a set of users denoted as $\mathcal{U}$, and a set of items denoted as $\mathcal{I}$. Users may have selection behaviors on some items, where $\mathcal{I}^u$ denotes the set of items selected by user $u$. Each item $i$ is associated with an image describing its visual information, and belongs to a specific category $l_i$. The whole set of categories is denoted as $\mathcal{L}$.

Moreover, users' selections on items are associated with timestamps. For each user $u$, corresponding categories form a sequence $\mathcal{L}^u = \{l_{1}^u,l_{2}^u,..., l_{t}^u,... \}$, where $l_t^u$ denotes the category of the item selected by user $u$ at time step $t$. At time step $t$ of the selection sequence of user $u$, there are input context $c_{I,t}^u$, i.e., the situation where the user conducts behavior, and transition context $c_{T,t}^u$, i.e., the time interval between the current timestamp and the previous timestamp.

\subsection{DeepStyle}

Conventional methods in visual recommendation are mostly focusing on modeling items in a common visual feature space. This may fail to capture different styles of items. As shown in Figure \ref{fig:origin}, items with similar styles may be not similar in the visual space at all. And categorical information is dominant in the common visual space. Thus, in visual recommendation, it is vital to eliminate characteristics of categories from representations of items. Accordingly, we propose the DeepStyle method for learning items' style features and users' preferences.

\begin{figure}[!tb]
\centering
\includegraphics[width=1\linewidth]{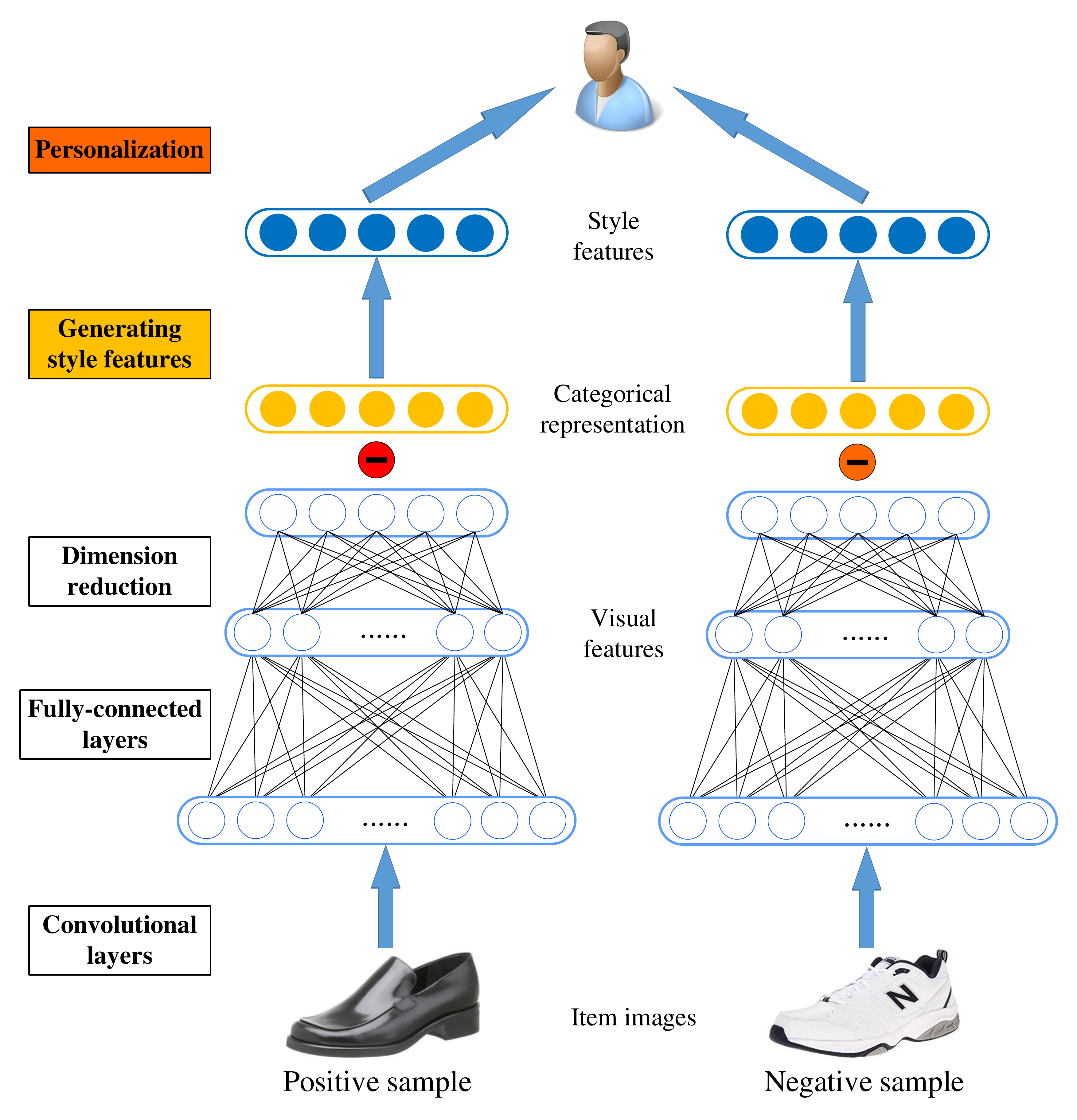}
\caption{The illustration of DeepStyle for learning styles of items and preferences of users. In DeepStyle, images of items are feeded into a deep CNN model. For each item, on the output layer of CNN, we subtract a latent representation of the corresponding category from the visual features generated by CNN, and obtain style features of items. The CNN architecture used here is the Caffe reference model \cite{mcauley2015inferring,mcauley2015image}, which consists of $5$ convolutional layers followed by $3$ fully-connected layers and is pre-trained on $1.2$ million ImageNet images. Finally, the style features are incorporated into a BPR framework \cite{rendle2009bpr} for personalized recommendation.}
\label{fig:ds}
\end{figure}

First, for each item $i$, we feed the corresponding image into a deep CNN model, as shown in Figure \ref{fig:ds}. Following several representative works \cite{he2016sherlock,he2016ups,he2016vbpr,mcauley2015image} in visual recommendation, the CNN model applied is the Caffe reference model \cite{mcauley2015inferring,mcauley2015image}. It consists of $5$ convolutional layers followed by $3$ fully-connected layers. The model is pre-trained on 1.2 million ImageNet images\footnote{http://image-net.org/}, for capturing some common visual concepts. On the output layer of the CNN model, there is a $4096$ dimensional visual feature vector denoted as ${{\mathbf{v}}_i} \in {\mathbb{R}^{4096}}$.

Then, to obtain style features, according to Equation \ref{eq:ass_style}, we subtract items' latent categorical representations from visual features generated by CNN. For item $i$, we can calculate its style features as
\begin{equation} \label{eq:style}
{{\mathbf{s}}_i} = {\mathbf{E}}{{\mathbf{v}}_i} - {{\mathbf{l}}_i}~,
\end{equation}
where ${{\mathbf{s}}_i} \in {\mathbb{R}^{d}}$ denotes the style feature of item $i$, ${{\mathbf{l}}_i} \in {\mathbb{R}^{d}}$ denotes the categorical representation of the corresponding category $l_i$, ${{\mathbf{E}}} \in {\mathbb{R}^{d \times 4096}}$ is a matrix for transferring visual features to lower dimensionality on the top layer, and $d$ is the dimensionality of learned representations.

Furthermore, similar with VBPR \cite{he2016vbpr}, we incorporate style features in the BPR \cite{rendle2009bpr} framework, which is the state-of-the-art method for modeling implicit feedbacks, for sensing preferences of users. The prediction of user $u$ on item $i$ can be made as
\begin{equation} \label{eq:style_predict}
{{\hat y}_{u,i}} = {\left( {{{\mathbf{p}}_{u}}} \right)^T}\left( {{{\mathbf{s}}_i}{ + }{{\mathbf{q}}_i}} \right)~,
\end{equation}
where ${{\mathbf{p}}_u} \in {\mathbb{R}^{d}}$ denotes the latent representation of user $u$, and ${{\mathbf{q}}_i} \in {\mathbb{R}^{d}}$ denotes the latent representation of item $i$. For user $u$, with an arbitrary negative sample $i'$, the model needs to fit
\begin{equation} \label{eq:style_dayu}
{{\hat y}_{u,i}} > {{\hat y}_{u,i'}}~,
\end{equation}
where $i$ is a positive item that $i \in {\mathcal{I}^u}$, and $i'$ is a negative item that $i' \notin {\mathcal{I}^u}$. Then, in the BPR framework, we need to maximize the following probability
\begin{equation} \label{eq:style_pro}
p\left( {u,i > i'} \right) = g\left( {{{\hat y}_{u,i}} - {{\hat y}_{u,i'}}} \right)~,
\end{equation}
where the activation function $f(x)$ is usually chosen as
\begin{equation} \label{eq:style_act}
g\left( x \right) = \frac{1}{{1 + {e^{ - x}}}}~.
\end{equation}
Incorporating the negative log likelihood, we can minimize the following objective function equivalently
\begin{equation} \label{eq:style_obj}
{J_p} = \sum\limits_{u,i} {\ln \left( {1 + {e^{ - \left( {{{\hat y}_{u,i}} - {{\hat y}_{u,i'}}} \right)}}} \right)}  + \frac{{{\lambda _p}}}{2}{\left\| {{{\mathbf{\theta }}_p}} \right\|^2}~,
\end{equation}
where ${\mathbf{\theta }}_p$ denotes all the parameters to be estimated in DeepStyle, and $\lambda _p$ is a hyper-parameter to control the power of regularization. Then, the derivations of $J_p$ with respect to all the parameters in DeepStyle can be calculated, and we can employ Stochastic Gradient Descent (SGD) to estimate the model parameters. The training procedure consists of two parts: the training of the deep CNN layers which generate visual features, and the training of the BPR layer which learns style features. These two parts of training are done alternately. This process is repeated iteratively until the convergence is achieved.

\subsection{CA-GRU}

\begin{figure}[!tb]
\centering
\includegraphics[width=1\linewidth]{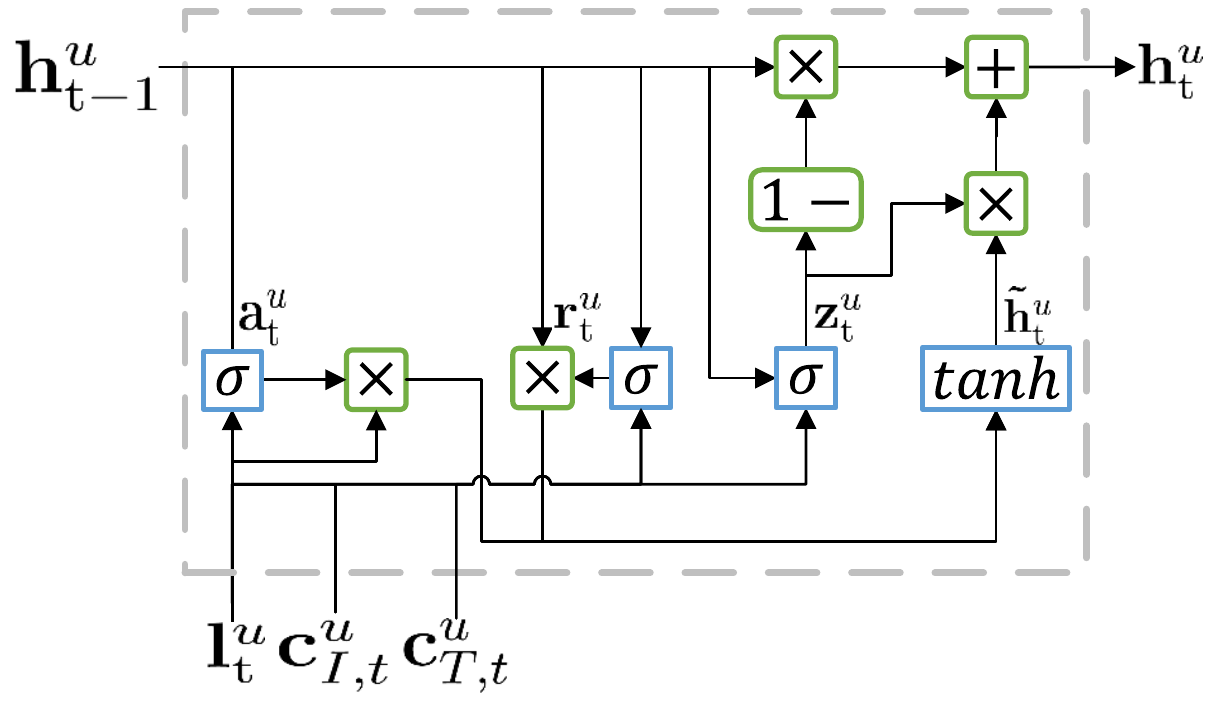}
\caption{The illustration of CA-GRU for learning users' demands. In CA-GRU, update gate, activation gate and reset gate are adjusted to variety of input contexts and transition contexts.}
\label{fig:ca}
\end{figure}

After proposing DeepStyle for learning users' preferences on different styles of items, we need to predict users' demands on different categories of items. Usually, demands are temporal. It depends on external situations, e.g., weather and month, and what a user has bought before. This make demand prediction a problem of recommendation with both contextual and sequential information. In our previous work \cite{liu2016context}, we address the problem of context-aware sequential recommendation, and propose a CA-RNN method based on the RNN architecture. CA-RNN adjusts matrices in the RNN formulation to various contexts. However, CA-RNN has its own drawbacks: 1) models constructed on conventional RNN architecture may occur the vanishing or exploding gradients problem \cite{bengio1994learning}; 2) assigning each context with a matrix requires too many parameters to be learned. Accordingly, we propose a CA-GRU method based on the GRU architecture.

First, we introduce the conventional GRU architecture \cite{chung2014empirical,chung2015gated}. For modeling our problem, the formulation of GRU is
\begin{equation} \label{eq:GRU}
\begin{array}{l}
{\mathbf{z}}_{{t}}^u = \sigma \left( {{{\mathbf{W}}_z}{\mathbf{l}}_{{t}}^u + {{\mathbf{M}}_z}{\mathbf{h}}_{{{t - 1}}}^u} \right)\\
{\mathbf{r}}_{{t}}^u = \sigma \left( {{{\mathbf{W}}_r}{\mathbf{l}}_{{t}}^u + {{\mathbf{M}}_r}{\mathbf{h}}_{{{t - 1}}}^u} \right)\\
{\mathbf{\tilde h}}_{{t}}^u = tanh\left( {{{\mathbf{W}}_h}{\mathbf{l}}_{{t}}^u + {{\mathbf{M}}_h}\left( {{\mathbf{h}}_{{{t - 1}}}^u \cdot {\mathbf{r}}_{{t}}^u} \right)} \right)\\
{\mathbf{h}}_{{t}}^u = \left( {1 - {\mathbf{z}}_{{t}}^u} \right) \cdot {\mathbf{h}}_{{{t - 1}}}^u + {\mathbf{z}}_{{t}}^u \cdot {{{\mathbf{\tilde h}}}_t}
\end{array}~,
\end{equation}
where ${{\mathbf{h}}_{{t}}^u} \in {\mathbb{R}^{d}}$ denotes the hidden state of user $u$ at time step $t$, ${{\mathbf{l}}_{{t}}^u} \in {\mathbb{R}^{d}}$ denotes the representation of the category $l_t^u$, ${{\mathbf{z}}_{{t}}^u} \in {\mathbb{R}^{d}}$ and ${{\mathbf{r}}_{{t}}^u} \in {\mathbb{R}^{d}}$ are the update gate and the reset gate respectively, $\mathbf{W}_*$ and $\mathbf{M}_*$ are all $d \times d$ dimensional matrices. With the gating operations, GRU can relieve the vanishing or exploding gradients problem to a certain extent.

According to the definition in \cite{liu2016context}, there are two types of contexts in sequences: input contexts and transition contexts. Input contexts are external contexts under which users conduct behaviors. Such contexts usually include location (home or working place), time (weekdays or weekends, morning or evening), weather (sunny or rainy), etc. Transition contexts denote time intervals between adjacent behaviors. It captures context-adaptive transition effects from past behaviors to future behaviors with different time intervals. That is to say, input and transition contexts affect current input elements and propagated previous hidden states respectively. Thus, the formulation in CA-GRU can be adjusted as
\begin{equation} \label{eq:CA-GRU}
\begin{array}{l}
{\mathbf{z}}_{{t}}^u = \sigma \left( {{{\mathbf{W}}_z}{\mathbf{l}}_{{t}}^u + {{\mathbf{M}}_z}{\mathbf{h}}_{{{t - 1}}}^u + {{\mathbf{I}}_z}{\mathbf{c}}_{I,t}^u + {{\mathbf{T}}_z}{\mathbf{c}}_{T,t}^u} \right)\\
{\mathbf{a}}_{{t}}^u = \sigma \left( {{{\mathbf{W}}_a}{\mathbf{l}}_{{t}}^u + {{\mathbf{M}}_a}{\mathbf{h}}_{{{t - 1}}}^u + {{\mathbf{I}}_a}{\mathbf{c}}_{I,t}^u} \right)\\
{\mathbf{r}}_{{t}}^u = \sigma \left( {{{\mathbf{W}}_r}{\mathbf{l}}_{{t}}^u + {{\mathbf{M}}_r}{\mathbf{h}}_{{{t - 1}}}^u + {{\mathbf{T}}_r}{\mathbf{c}}_{T,t}^u} \right)\\
{\mathbf{\tilde h}}_{{t}}^u = tanh\left( {{{\mathbf{W}}_h}\left( {{\mathbf{l}}_{{t}}^u \cdot {\mathbf{a}}_{{t}}^u} \right) + {{\mathbf{M}}_h}\left( {{\mathbf{h}}_{{{t - 1}}}^u \cdot {\mathbf{r}}_{{t}}^u} \right)} \right)\\
{\mathbf{h}}_{{t}}^u = \left( {1 - {\mathbf{z}}_{{t}}^u} \right) \cdot {\mathbf{h}}_{{{t - 1}}}^u + {\mathbf{z}}_{{t}}^u \cdot {{{\mathbf{\tilde h}}}_t}
\end{array}~,
\end{equation}
where ${{\mathbf{c}}_{I,t}^u} \in {\mathbb{R}^{d}}$ and ${{\mathbf{c}}_{T,t}^u} \in {\mathbb{R}^{d}}$ denote vector representations of input context $c_{I,t}^u$ and transition context $c_{T,t}^u$ respectively, $\mathbf{I}_*$ and $\mathbf{T}_*$ are all $d \times d$ dimensional matrices. Considering the update gate controls the updating weights between current states and previous states, we incorporate ${\mathbf{z}}_{{t}}^u$ with both input and transition contexts. Because the reset gate resets the propagated signals from previous states, we incorporate ${\mathbf{r}}_{{t}}^u$ with only transition contexts. Moreover, we add an activation gate ${{\mathbf{a}}_{{t}}^u} \in {\mathbb{R}^{d}}$ incorporating input contexts for modeling the activating operation on input elements ${{\mathbf{l}}_{{t}}^u}$. With context-aware gate vectors, CA-GRU significantly reduces the number of parameters comparing with CA-RNN.

When making prediction at time step $t+1$, despite operations discussed above, current contexts $c_{I,t+1}^u$ and $c_{T,t+1}^u$ should also be considered. Then, the prediction of user $u$ at time step $t+1$ on category $j$ can be made as
\begin{equation} \label{eq:CAGRU_predict}
\begin{array}{l}
{\mathbf{a}'}_{t + 1}^u = \sigma \left( {{\mathbf{I}'}{_a}{\mathbf{c}}_{I,t + 1}^u} \right)\\
{\mathbf{r}'}_{t + 1}^u = \sigma \left( {{\mathbf{T}'}{_r}{\mathbf{c}}_{T,t + 1}^u} \right)\\
{{\hat y}_{u,t + 1,j}} = {\left( {{\mathbf{h}}_{{t}}^u \cdot {\mathbf{r}'}_{t + 1}^u} \right)^T}\left( {{{\mathbf{l}}_j} \cdot {\mathbf{a}'}_{t + 1}^u} \right)
\end{array}~,
\end{equation}
where ${\mathbf{a}'}_{t + 1}^u$ and ${\mathbf{r}'}_{t + 1}^u$ are the activation gate and the reset gate at the prediction scenario respectively, and as same as $\mathbf{I}_*$ and $\mathbf{T}_*$, $\mathbf{I}'_*$ and $\mathbf{T}'_*$ are all $d \times d$ dimensional matrices.

Furthermore, we incorporate the cross-entropy for learning of CA-GRU. We need to minimize the following objective function
\begin{equation} \label{eq:CAGRU_obj}
{J_d} =  - \sum\limits_{u,t} {\frac{1}{{\left| \mathcal{L} \right|}}\sum\limits_{j \in \mathcal{L}} {{y_{u,t + 1,j}}\ln \left( {{{\hat y}_{u,t + 1,j}}} \right)} }  + \frac{{{\lambda _d}}}{2}{\left\| {{{\mathbf{\theta }}_d}} \right\|^2}~,
\end{equation}
where ${\mathbf{\theta }}_d$ denotes all the parameters to be learned in CA-GRU, and $\lambda _d$ is the regularization parameter. According to the objective function, parameters in CA-GRU can be estimated with the commonly-used BPTT \cite{rumelhart1988learning} and SGD. To be noted, CA-GRU can be pre-trained on DeepStyle. That us to say, ${{\mathbf{l}}_{{t}}^u}$ can be initialized with the categorical representations learned in DeepStyle. With this trick, CA-GRU is able to capture some visual characteristics and correlations.

\subsection{Summary of Proposed Methods}

\begin{figure}[!tb]
\centering
\includegraphics[width=1\linewidth]{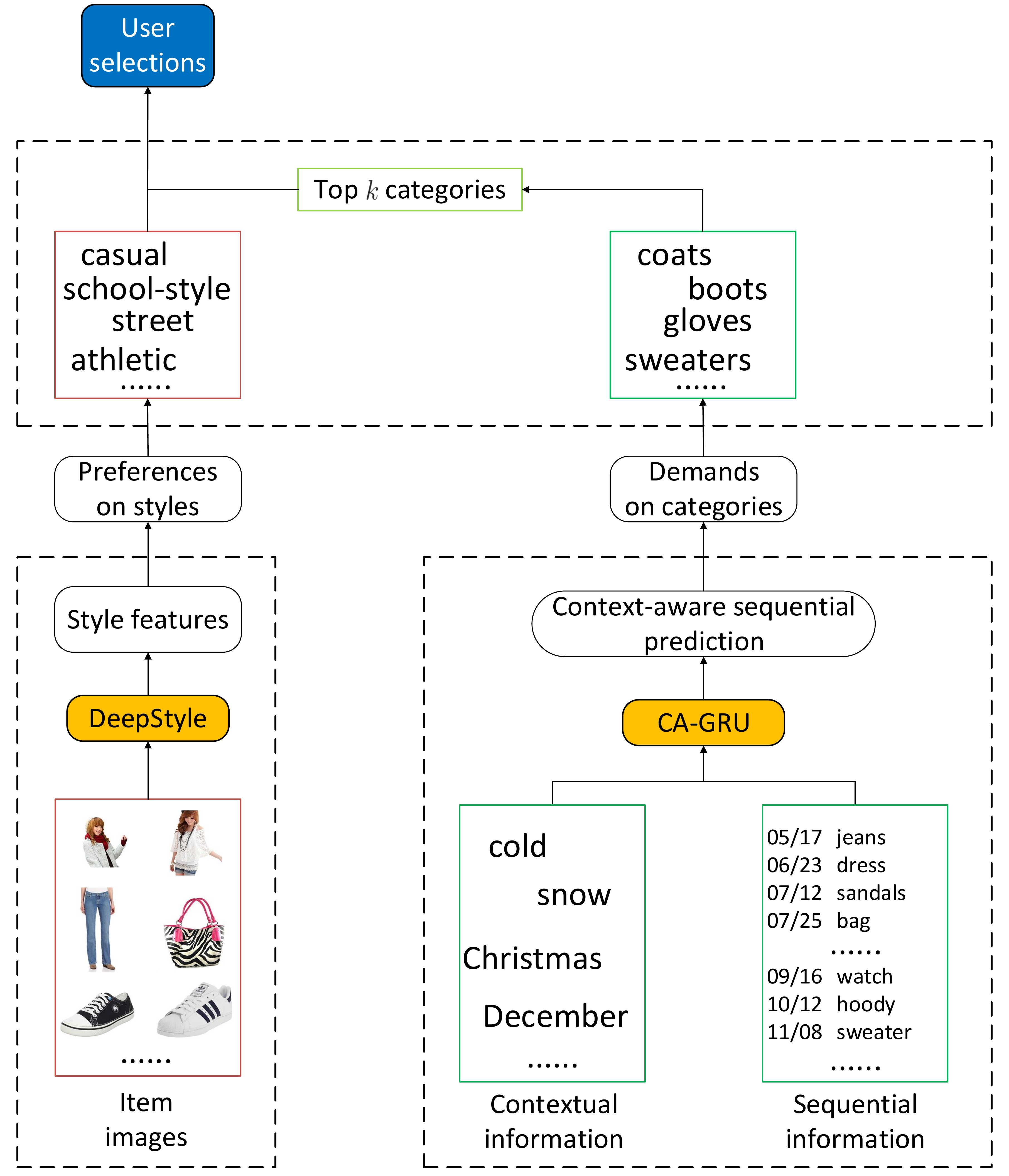}
\caption{The summary of our proposed methods, i.e., DeepStyle and CA-GRU. Based on visual information, DeepStyle learns style features of items and preferences of users, as shown in the bottom left square. Based on contextual information and sequential information, CA-GRU learns users' demands on categories, as shown in the bottom right square. Then, we can obtain top $k$ categories with largest probabilities predicted by CA-GRU. Among these categories, items matching users' preferences will be recommended, as shown in the top square.}
\label{fig:frame}
\end{figure}

As shown in Figure \ref{fig:frame}, based on visual information, DeepStyle learns users' preferences on styles. And based on contextual information and sequential information, CA-GRU learns users' demands on categories. Finally, we need to aggregate the prediction on preferences and demands for better predicting users' selections, and generate a ranked list of items for each user at each predicted time step. First, based on the predicted results generated by CA-GRU, we pick top $k$ categories with largest probabilities. Items in these top $k$ categories form the former part of the list, and items in other categories form the latter part of the list. Second, in either part of the list, items are ranked according to the matching degrees with the user's preferences measured by DeepStyle. Then, the final recommended results are generated. And the whole procedure of recommendation can be named as DeepStyle+CA-GRU for predicting users' selections on items in visual recommendation.

\section{Experiments}

In this section, we introduce our experiments to evaluate the effectiveness of our proposed methods. First, we introduce our experimental settings. Then, we give comparison among some state-of-the-art methods in different aspects. Finally, we demonstrate the visualization of clustering results measured by style features learned in DeepStyle.

\subsection{Experimental Settings}

\begin{table*}[tb!]
  \centering
  \caption{Performance comparison on predicting users' selections on items measured by AUC. The dimensionality is $d=10$ on both datasets. Numbers of top categories are $k=3$ and $k=5$ on Clothing and Home respectively. CA-GRU is pre-trained and initialized with DeepStyle.}
    \begin{tabular}{ccccccc}
    \toprule
    dataset & setting & BPR   & VBPR  & Sherlock & DeepStyle & DeepStyle+CA-GRU \\
    \midrule
    \multirow{2}[2]{*}{Clothing} & warm-start & 0.6183  & 0.7441  & 0.7758  & \textbf{0.7961} & \textbf{0.8075} \\
          & cold-start & 0.5037  & 0.6915  & 0.7167  & \textbf{0.7317} & \textbf{0.7599} \\
    \midrule
    \multirow{2}[2]{*}{Home} & warm-start & 0.5848  & 0.6845  & 0.7049  & \textbf{0.7155} & \textbf{0.7390} \\
          & cold-start & 0.5053  & 0.6140  & 0.6322  & \textbf{0.6396} & \textbf{0.6711} \\
    \bottomrule
    \end{tabular}%
  \label{tab:selection}%
\end{table*}%

Our experiments are conducted on two subsets of the Amazon dataset \cite{mcauley2015inferring,mcauley2015image}. In particular, we adopt the ``Clothing, Shoes and Jewelry" subset and the ``Home and Kitchen" subset, which are named as the \textbf{Clothing} dataset and the \textbf{Home} dataset for short. The reason we choose these two datasets is that, visual features are important in buying things such as clothes, shoes, jewelries, house decorations and so on. Especially, visual features have been proven to be useful in cloth recommendation \cite{he2016sherlock,he2016ups,he2016vbpr,mcauley2015image}. The Clothing dataset consists of $74$ categories, e.g., jeans, pants, shoes, shirts and dresses. The home dataset contains $86$ categories, e.g., sheets, furniture, pillows and cups.

In our experiments, we empirically set the regulation parameters as $\lambda_p = 0.01$ and $\lambda_d = 0.01$, and the learning rate for SGD is set to be $0.01$. For each dataset, we use $80\%$ instances for training, and remaining $20\%$ instances for testing. Moreover, we remove users with less than 5 records and more than 100 records. There are two types of evaluation settings on both datasets during the testing procedure: \textbf{warm-start} and \textbf{cold-start}. The former focuses on measuring the overall ranking performance, while the latter captures the capability to recommend cold-start items, i.e., items with less than $5$ records during training, in the system.

According to timestamps associated in the two datasets, similar with the operations in \cite{liu2016context}, we extract input contexts and transition contexts for CA-GRU, CA-RNN and other context-aware methods. First, we can extract two kinds of contexts: seven days in a week and twelve months in a year. So, there are totally $84$ input context values. Second, we can extract time intervals between adjacent behaviors in sequences as transition contexts. Time intervals in both datsets are empirically discretized to several time bins, where thresholds are one day, two days, three days, one week, half a month, one month, three months, half a year and one year. So, there are totally $10$ transition context values.

Then, following some previous works \cite{he2016vbpr,rendle2009bpr}, for evaluating the performance of all the methods, we apply the Area Under the ROC Curve (\textbf{AUC}) metric:
\begin{small}
\begin{displaymath}
AUC = \frac{1}{{\left| \mathcal{U} \right|}}\sum\limits_{u \in \mathcal{U}} {\frac{1}{{\left| {set\left( {i \in {\mathcal{I}^u},i' \notin {\mathcal{I}^u}} \right)} \right|}}\sum\limits_{i \in {\mathcal{I}^u},i' \notin {\mathcal{I}^u}} {\delta \left( {{p_{u,i}} > {p_{u,i'}}} \right)} }~,
\end{displaymath}
\end{small}
where $\delta \left( . \right)$ is the Dirac delta function, which outputs $1$ when the condition is met, and $0$ otherwise. The larger the AUC value, the better the performance.

Moreover, to illustrate the effectiveness of our proposed methods, several aspects of methods are compared:

(1) To investigate the performance on predicting users' preferences on styles, some state-of-the-art methods in visual recommendation are compared: \textbf{BPR} \cite{rendle2009bpr}, \textbf{VBPR} \cite{he2016vbpr} and \textbf{Sherlock} \cite{he2016sherlock}. BPR is a widely-used method for modeling implicit feedbacks. Based on BPR, VBPR incorporates visual features of items. Sherlock extends VBPR, and takes categorical effects on styles into consideration. As in \cite{he2016sherlock,he2016vbpr}, visual features used in VBPR and Sherlock are CNN features extracted from the Caffe reference model \cite{jia2014caffe,krizhevsky2012imagenet}.

(2) To investigate the performance on predicting users' demands on categories, several methods in both context-aware recommendation and sequential recommendation are compared. Context-aware methods consist of \textbf{FM} \cite{rendle2011fast} and \textbf{COT} \cite{liu2015cot}, while sequential methods include \textbf{FPMC} \cite{rendle2010factorizing}, \textbf{HRM} \cite{wang2015learning}, \textbf{RNN} \cite{yu2016dream} and \textbf{GRU} \cite{hidasi2016session}. Moreover, \textbf{CA-RNN} \cite{liu2016context}, which can model contextual and sequential information simultaneously, is also compared.

\subsection{Evaluation of DeepStyle}

\begin{figure}[tb!]
\centering
\hspace{-6mm}
\subfigure[Clothing.]{
\begin{minipage}[b]{0.25\textwidth}
\centering
\includegraphics[width=1\textwidth]{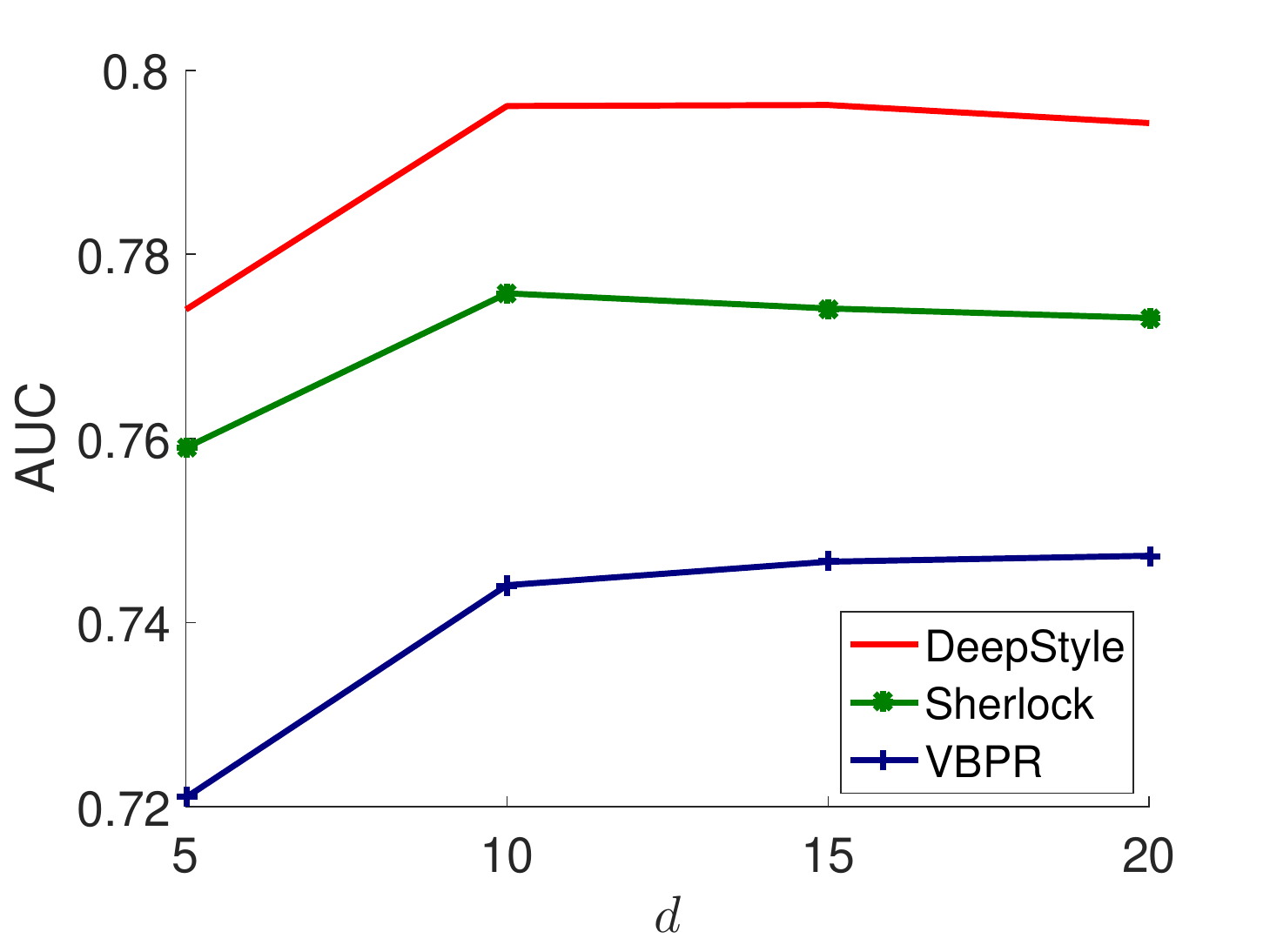}
\label{dimen1}
\end{minipage}
}
\hspace{-6mm}
\subfigure[Home.]{
\begin{minipage}[b]{0.25\textwidth}
\centering
\includegraphics[width=1\textwidth]{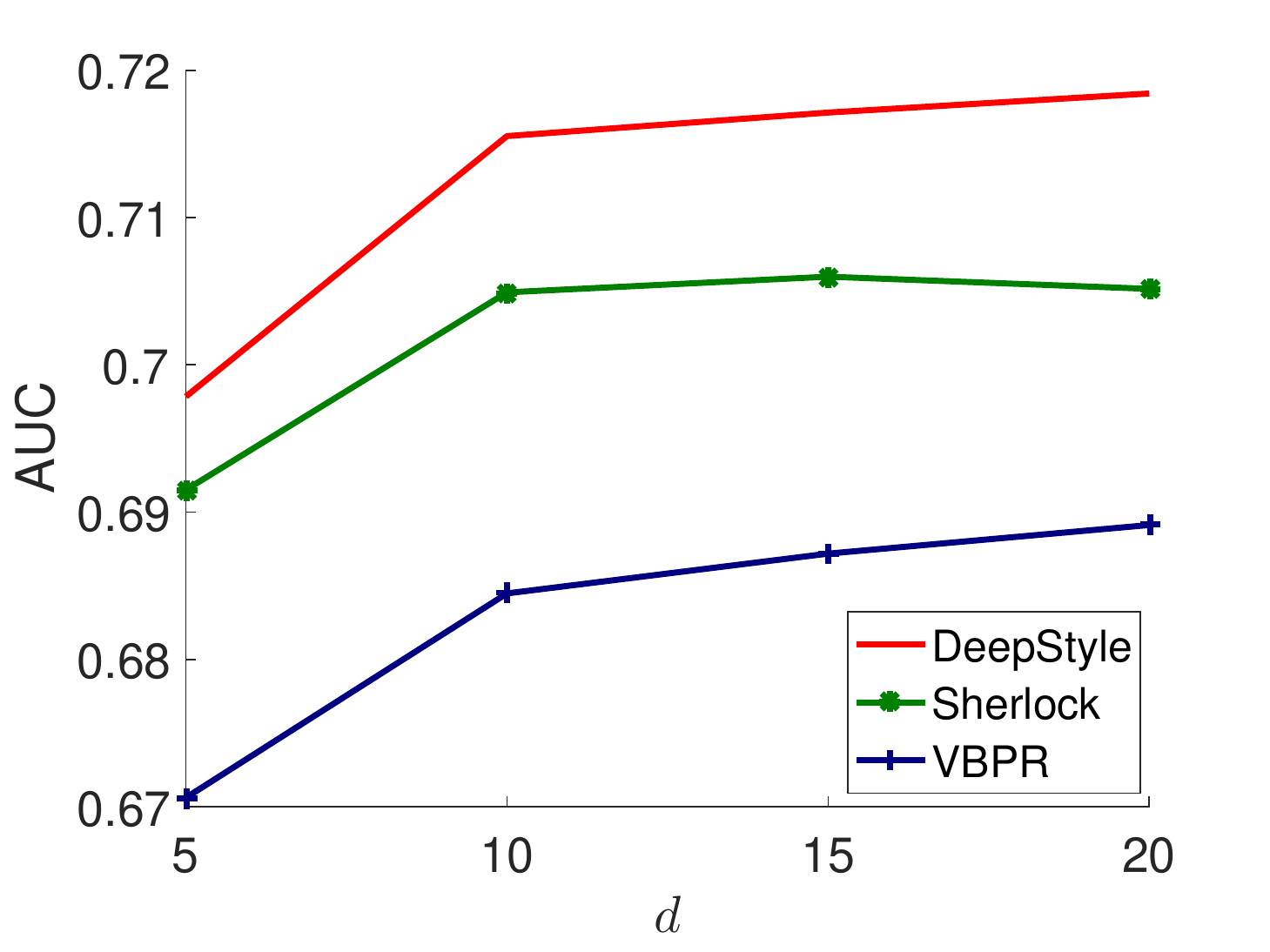}
\label{dimen2}
\end{minipage}
}
\caption{Performance of DeepStyle, Sherlock and VBPR with varying dimensionality $d=[5,10,15,20]$ measured by AUC.}
\label{fig:dim}
\end{figure}

\begin{table*}[tb!]
  \centering
  \caption{Performance comparison on context-aware sequential recommendation, i.e., predicting users' demands on categories, measured by AUC. The dimensionality is $d=10$ on both datasets.}
    \begin{tabular}{c|cc|cccc|ccc}
    \toprule
    \multirow{2}[4]{*}{dataset} & \multicolumn{2}{c|}{context-aware} & \multicolumn{4}{c|}{sequential} & \multicolumn{3}{c}{context-aware+sequential} \\
\cmidrule{2-10}          & FM    & COT   & FPMC  & HRM   & RNN   & GRU   & CA-RNN & CA-GRU & CA-GRU(pre-trained) \\
    \midrule
    Clothing & 0.6704  & 0.6716  & 0.6928  & 0.6973  & 0.7169  & 0.7422  & 0.7483  & \textbf{0.7582} & \textbf{0.7602} \\
    Home  & 0.6692  & 0.6711  & 0.6789  & 0.6846  & 0.7021  & 0.7268  & 0.7241  & \textbf{0.7351} & \textbf{0.7385} \\
    \bottomrule
    \end{tabular}%
  \label{tab:demand}%
\end{table*}%

Table \ref{tab:selection} illustrates the performance comparison among DeepStyle, Sherlock, VBPR and BPR under warm-start and cold-start settings. And the dimensionality is $d=10$. We can clearly observe that, methods incorporating visual features can outperform the baseline method BPR with relatively large advantages on both datasets. The advantages comparing with BPR are even larger under the cold-start setting, which indicates that visual features can model properties of cold-start items when observations are not enough, and promote the cold-start evaluation. Moreover, methods modeling categorical effects on styles of items, i.e., Sherlock and DeepStyle, have better performance than VBPR on both datasets under both settings. This shows it is vital to take categorical information into consideration for modeling styles of items. And Sherlock becomes the best one among all the compared methods in visual recommendation. It is obvious that our proposed DeepStyle method outperforms all the compared methods. Comparing with Sherlock, DeepStyle improves AUC values by $2.3\%$ and $1.1\%$ on Clothing and Home respectively under the warm-start setting, and $1.5\%$ and $0.7\%$ under the cold-start setting. These improvements indicates the superiority of DeepStyle for learning style features of items and preferences of users.

Moreover, we illustrate the dimensionality sensitivity of DeepStyle, Sherlock and VBPR in Figure \ref{fig:dim}. DeepStyle can consistently outperform Sherlock and VBPR. On both datasets, the performance of DeepStyle stays stable after $d=10$. This indicates DeepStyle is not sensitive with the dimensionality, and the performance with $d=10$ is reported in the rest of our experiments.

\begin{figure}[tb!]
\centering
\hspace{-6mm}
\subfigure[Clothing.]{
\begin{minipage}[b]{0.25\textwidth}
\centering
\includegraphics[width=1\textwidth]{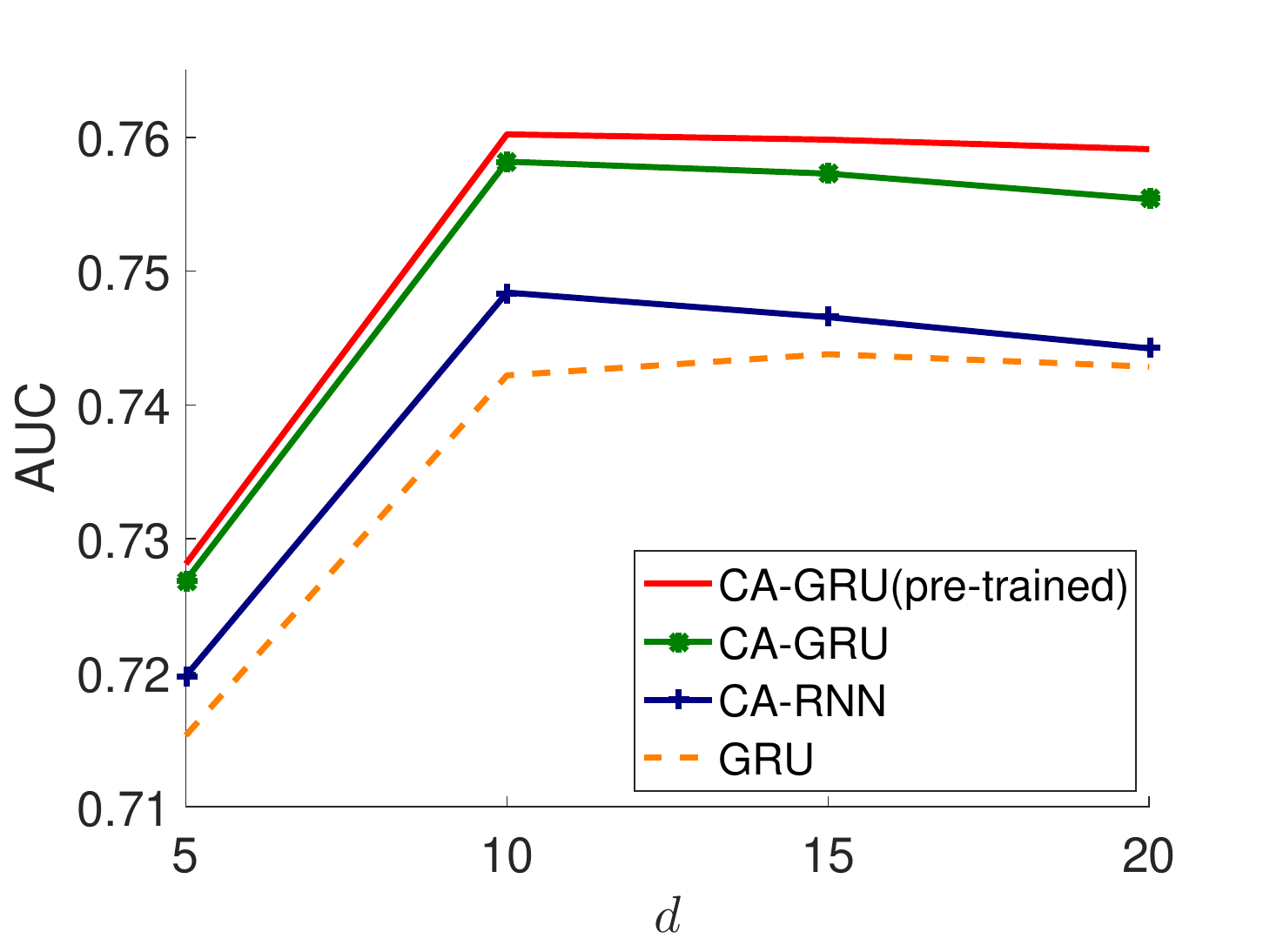}
\label{dimen1}
\end{minipage}
}
\hspace{-6mm}
\subfigure[Home.]{
\begin{minipage}[b]{0.25\textwidth}
\centering
\includegraphics[width=1\textwidth]{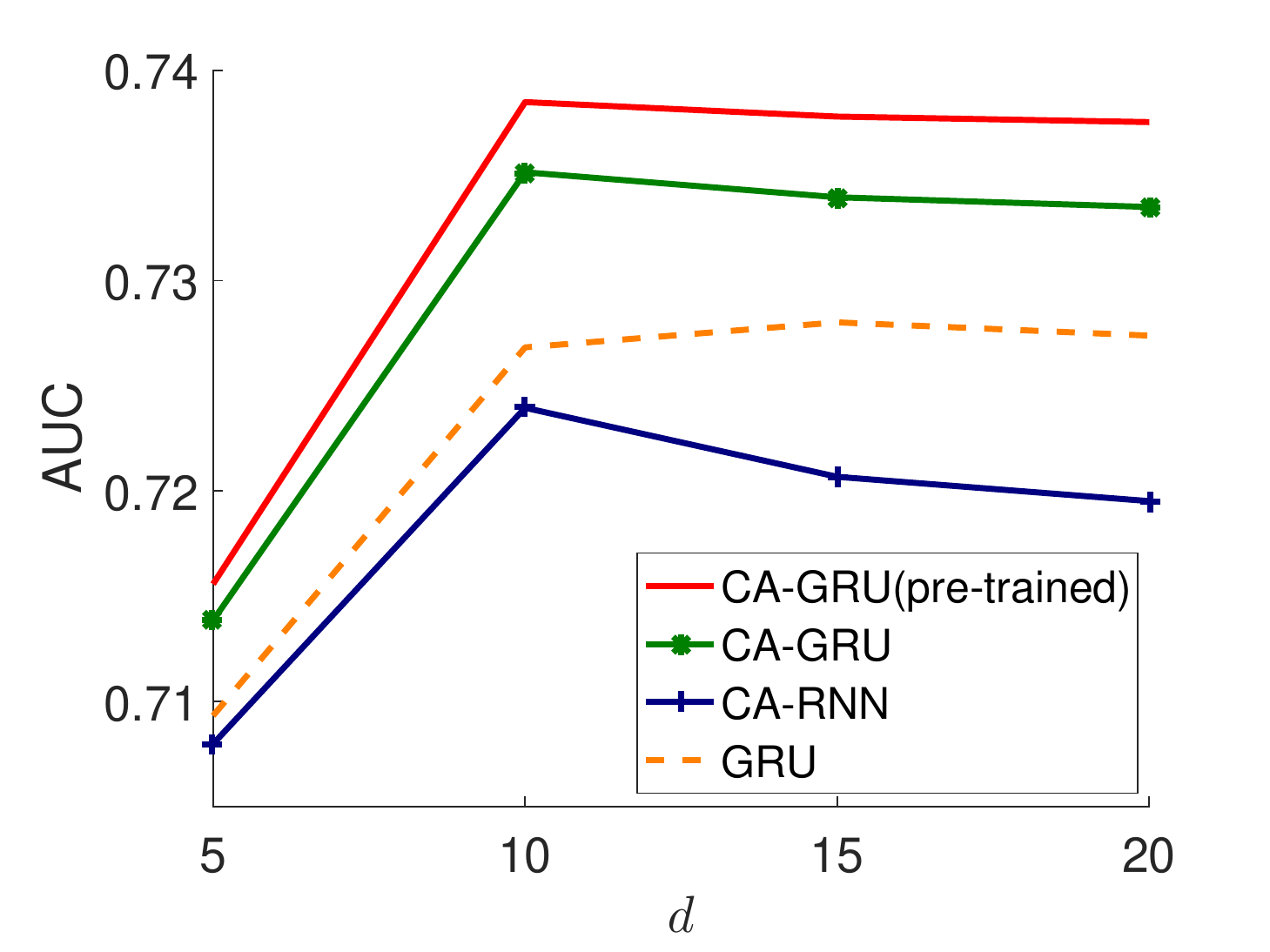}
\label{dimen2}
\end{minipage}
}
\caption{Performance of CA-GRU, CA-RNN and GRU with varying dimensionality $d=[5,10,15,20]$ measured by AUC.}
\label{fig:cd}
\end{figure}

\subsection{Evaluation of CA-GRU}

To investigate the performance of CA-GRU in context-aware sequential recommendation, its comparison with CA-RNN and some state-of-the-art methods in context-aware and sequential recommendation is shown in Table \ref{tab:demand}. The dimensionality is $d=10$ on both datasets. Comparing with context-aware methods, i.e., FM and COT, sequential methods, i.e., FPMC, HRM, RNN and GRU, have better performance. This show that, sequential information is usually important than contextual information in predicting users' demands. GRU can significantly outperform RNN, which indicates the advantage of the gating operations for solving the the vanishing or exploding gradients problem. GRU and CA-RNN have similar performance, where CA-RNN performs better on Clothing, while GRU performs better on Home. This may because that, sequences in the Home dataset are longer, and the performance benefits from the gating operations. Moreover, it is obvious that CA-GRU can outperform all the compared methods on both datasets. And pre-trained with DeepStyle, the performance improves slightly furthermore. Comparing with CA-RNN, CA-GRU improves AUC values by $1.2\%$ and $1.4\%$ on Clothing and Home respectively. This shows the advantages of CA-GRU in context-aware sequential recommendation and predicting users' demands on different categories of items.

Performance of CA-GRU, CA-RNN and GRU with varying dimensionality is also shown in Figure \ref{fig:cd}. CA-GRU can clearly outperform CA-RNN and GRU with all the dimensionality on both datasets. And the performance of CA-GRU stays stable after $d=10$. Accordingly, in the rest of our experiments, CA-GRU is reported with the dimensionality $d=10$. Moreover, CA-RNN tends to overfit the data when dimensionality is high, which proves context-aware matrices in CA-RNN require too many parameters to be learned.

\begin{figure}[tb!]
\centering
\hspace{-6mm}
\subfigure[Clothing.]{
\begin{minipage}[b]{0.25\textwidth}
\centering
\includegraphics[width=1\textwidth]{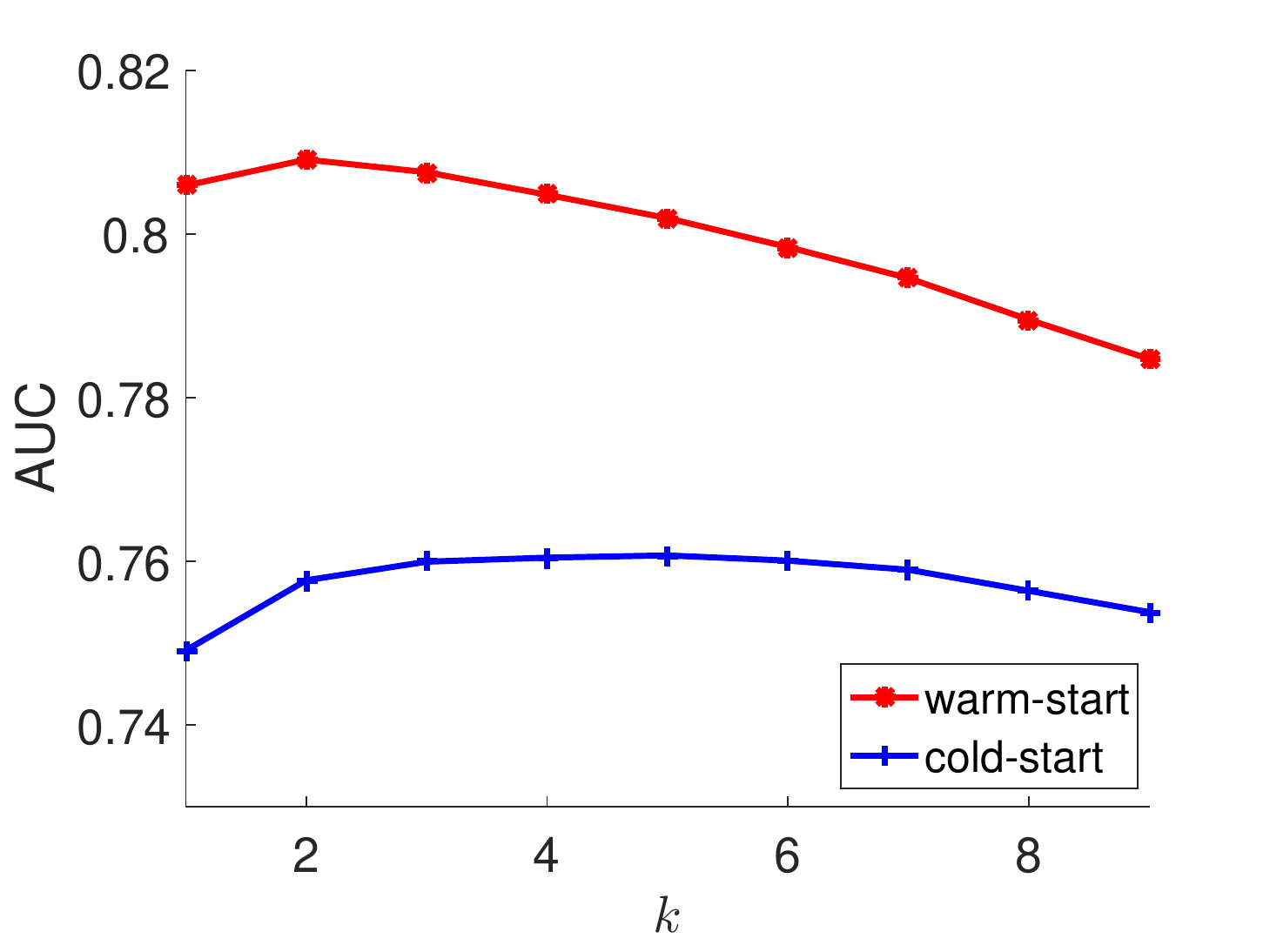}
\label{dimen1}
\end{minipage}
}
\hspace{-6mm}
\subfigure[Home.]{
\begin{minipage}[b]{0.25\textwidth}
\centering
\includegraphics[width=1\textwidth]{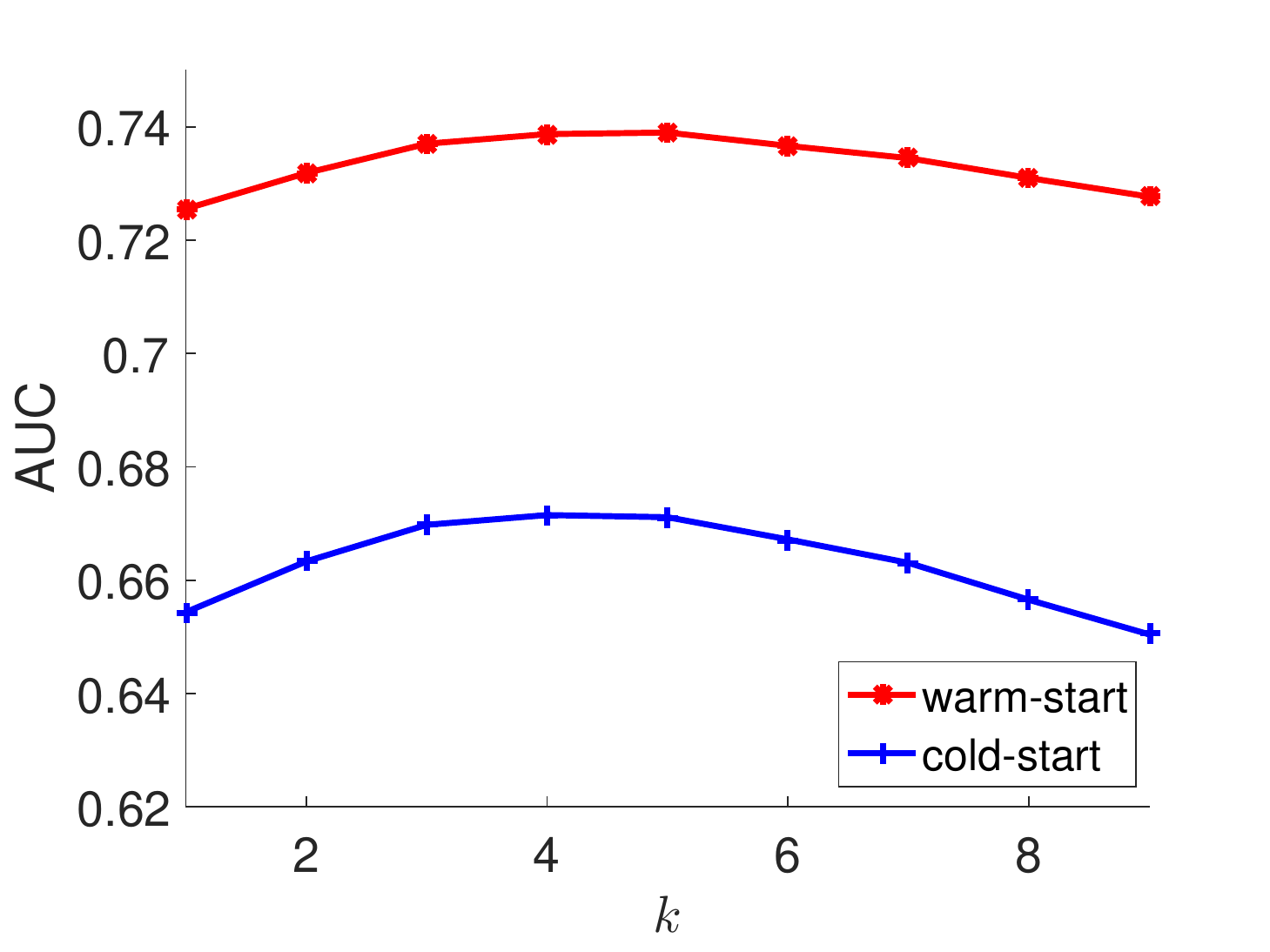}
\label{dimen2}
\end{minipage}
}
\caption{Performance of DeepStyle+CA-GRU with varying numbers of top categories $k=[1,2,3,4,5,6,7,8,9]$ measured by AUC. The dimensionality is $d=10$ on both datasets. CA-GRU is pre-trained and initialized with DeepStyle.}
\label{fig:top}
\end{figure}

\subsection{Evaluation of DeepStyle+CA-GRU}

\begin{figure*}[!tb]
\centering
\includegraphics[width=1\linewidth]{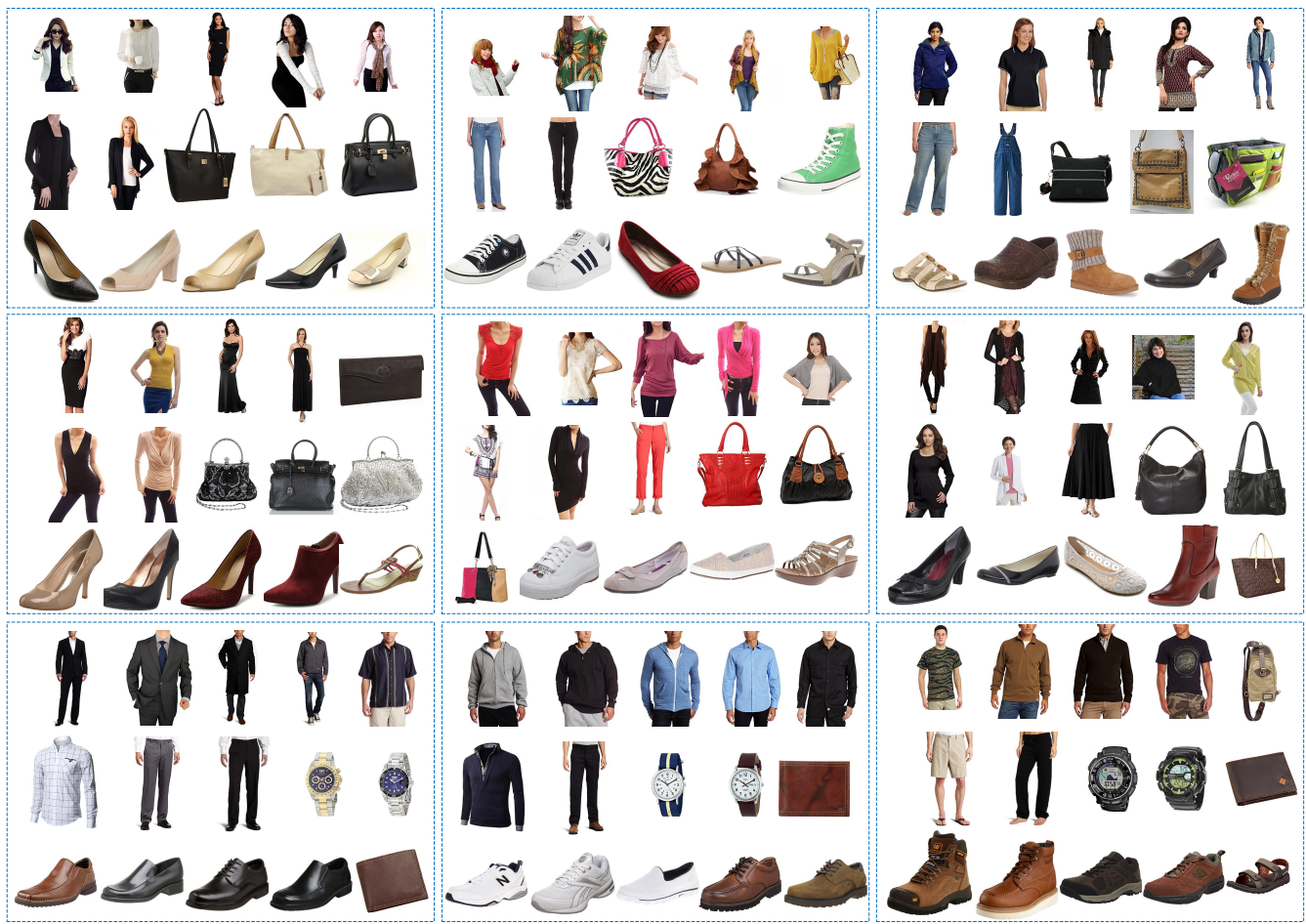}
\caption{Visualization of part of the clustering results of items in the Clothing dataset measured by the learned style features in DeepStyle. Items in one square belong to the same cluster. We can observe that, one category of items are assigned to different clusters. Male and female items are distinguished. And each cluster covers a distinct style of clothing.}
\label{fig:vis}
\end{figure*}

For better predicting users' selections, we can aggregate the prediction of DeepStyle and CA-GRU. Performance of DeepStyle+CA-GRU with varying numbers of top categories $k$ is illustrated in Figure \ref{fig:top}. CA-GRU is pre-trained with DeepStyle, and the dimensionality is $d=10$ for both DeepStyle and CA-GRU. We can clearly observe that, the performance of DeepStyle+CA-GRU is stable in a large range of $k$ under both warm- and cold-start evaluations, which indicates our proposed methods are not very sensitive with the parameter $k$. Integrated the results of warm- and cold-start evaluations, we select $k=3$ and $k=5$ as best parameters for DeepStyle+CA-GRU in Clothing and Home respectively.

Performance of DeepStyle+CA-GRU with best parameters is also shown and compared in Table \ref{tab:selection}. On both datasets, on the basis of DeepStyle, the performance is further improved via aggregating the prediction on preferences and demands. Under the warm-start evaluation, comparing with Sherlock, DeepStyle+CA-GRU improves AUC values by $3.2\%$ and $2.5\%$ on Clothing and Home respectively. Under the cold-start evaluation, the improvements become $4.3\%$ and $3.9\%$ on Clothing and Home respectively. It is obvious that, improvements are larger under the cold-start setting. This is because that, predicting users' demands on categories faces no cold-start problem, i.e., there is always enough data for learning representation of each category during training. These significant improvements show the rationality and advancement of our proposed DeepStyle and CA-GRU in visual recommendation.

\subsection{Visualization}

Based on 10-dimensional style features learned in DeepStyle, items in the Clothing dataset are clustered into several distinct styles. The visualization of part of the clustering results is shown in Figure \ref{fig:vis}. It is obvious that, one category of items are assigned to different clusters, and different styles of items are distinguished. Female items are in the top two rows, and male items are in the bottom row. The left column covers formal and official styles of clothing, in which the middle square is closer to the banquet-style. Items in the middle column are mostly casual, school-style or street-style clothing for women and men. In the right column, items somehow belong to the old-style, and the middle square is more likely the clothing style of middle-aged women. Each cluster clearly covers a distinct style of clothing. To be noted, during the training of DeepStyle, there is absolutely no supervision on styles. Thus, our proposed method is able to automatically capture different styles of items, and promote the visual recommendation.

\section{Conclusions and Future Work}

In this paper, we propose two novel methods, DeepStyle and CA-GRU, for learning users' preferences and demands in visual recommendation. Based on the CNN architecture and the BPR framework, DeepStyle subtracts categorical characteristics from visual features of items, and thus obtains style features of items. CA-GRU incorporates context-aware gates in the GRU formulation for adjusting to different contexts, and promoting the modeling of context-aware sequential recommendation. Experimental results demonstrate the successful performance of our proposed methods in visual recommendation.

In the future, we will further investigate the following directions. First, we will investigate deeper visual characteristics of items, to obtain the corresponding high-level semantic information. Second, we are going to analyze the long-term evolution of style features of items and preferences of users. Third, we plan to enlarge the range of research, which may include investigating images posted on various social media.

\balance
\bibliographystyle{ACM-Reference-Format}
\bibliography{style}

\end{document}